\PassOptionsToPackage{unicode}{hyperref}
\PassOptionsToPackage{hyphens}{url}
\PassOptionsToPackage{dvipsnames,svgnames,x11names}{xcolor}
\documentclass[
  letterpaper,
  DIV=11,
  numbers=noendperiod]{scrartcl}

\usepackage{amsmath,amssymb}
\usepackage{lmodern}
\usepackage{iftex}
\ifPDFTeX
  \usepackage[T1]{fontenc}
  \usepackage[utf8]{inputenc}
  \usepackage{textcomp} 
\else 
  \usepackage{unicode-math}
  \defaultfontfeatures{Scale=MatchLowercase}
  \defaultfontfeatures[\rmfamily]{Ligatures=TeX,Scale=1}
\fi
\IfFileExists{upquote.sty}{\usepackage{upquote}}{}
\IfFileExists{microtype.sty}{
  \usepackage[]{microtype}
  \UseMicrotypeSet[protrusion]{basicmath} 
}{}
\makeatletter
\@ifundefined{KOMAClassName}{
  \IfFileExists{parskip.sty}{%
    \usepackage{parskip}
  }{
    \setlength{\parindent}{0pt}
    \setlength{\parskip}{6pt plus 2pt minus 1pt}}
}{
  \KOMAoptions{parskip=half}}
\makeatother
\usepackage{xcolor}
\ifx\paragraph\undefined\else
  \let\oldparagraph\paragraph
  \renewcommand{\paragraph}[1]{\oldparagraph{#1}\mbox{}}
\fi
\ifx\subparagraph\undefined\else
  \let\oldsubparagraph\subparagraph
  \renewcommand{\subparagraph}[1]{\oldsubparagraph{#1}\mbox{}}
\fi

\usepackage{longtable,booktabs,array}
\usepackage{calc} 
\usepackage{etoolbox}
\makeatletter
\patchcmd\longtable{\par}{\if@noskipsec\mbox{}\fi\par}{}{}
\makeatother
\IfFileExists{footnotehyper.sty}{\usepackage{footnotehyper}}{\usepackage{footnote}}
\makesavenoteenv{longtable}
\usepackage{graphicx}
\makeatletter
\def\maxwidth{\ifdim\Gin@nat@width>\linewidth\linewidth\else\Gin@nat@width\fi}
\def\maxheight{\ifdim\Gin@nat@height>\textheight\textheight\else\Gin@nat@height\fi}
\makeatother
\setkeys{Gin}{width=\maxwidth,height=\maxheight,keepaspectratio}
\makeatletter
\def\fps@figure{htbp}
\makeatother
\newlength{\cslhangindent}
\newlength{\csllabelwidth}
\newlength{\cslentryspacingunit} 
\newenvironment{CSLReferences}[2] 
 {
  \setlength{\parindent}{0pt}
  \ifodd #1
  \let\oldpar\par
  \def\par{\hangindent=\cslhangindent\oldpar}
  \fi
  \setlength{\parskip}{#2\cslentryspacingunit}
 }%
 {}
\usepackage{calc}

\KOMAoption{captions}{tableheading}
\makeatletter
\@ifpackageloaded{caption}{}{\usepackage{caption}}
\AtBeginDocument{%
\ifdefined\contentsname
  \renewcommand*\contentsname{Table of contents}
\else
  \newcommand\contentsname{Table of contents}
\fi
\ifdefined\listfigurename
  \renewcommand*\listfigurename{List of Figures}
\else
  \newcommand\listfigurename{List of Figures}
\fi
\ifdefined\listtablename
  \renewcommand*\listtablename{List of Tables}
\else
  \newcommand\listtablename{List of Tables}
\fi
\ifdefined\figurename
  \renewcommand*\figurename{Figure}
\else
  \newcommand\figurename{Figure}
\fi
\ifdefined\tablename
  \renewcommand*\tablename{Table}
\else
  \newcommand\tablename{Table}
\fi
}
\@ifpackageloaded{float}{}{\usepackage{float}}
\floatstyle{ruled}
\@ifundefined{c@chapter}{\newfloat{codelisting}{h}{lop}}{\newfloat{codelisting}{h}{lop}[chapter]}
\floatname{codelisting}{Listing}

\makeatother
\makeatletter
\@ifpackageloaded{caption}{}{\usepackage{caption}}
\@ifpackageloaded{subcaption}{}{\usepackage{subcaption}}
\makeatother
\makeatletter
\@ifpackageloaded{tcolorbox}{}{\usepackage[many]{tcolorbox}}
\makeatother
\makeatletter
\@ifundefined{shadecolor}{\definecolor{shadecolor}{rgb}{.97, .97, .97}}
\makeatother
\ifLuaTeX
  \usepackage{selnolig}  
\fi
\IfFileExists{bookmark.sty}{\usepackage{bookmark}}{\usepackage{hyperref}}
\IfFileExists{xurl.sty}{\usepackage{xurl}}{} 
\hypersetup{
  pdftitle={A Digital Twin for Geological Carbon Storage with Controlled Injectivity},
  pdfauthor={Abhinav Prakash Gahlot1*, Haoyun Li,1*, Ziyi Yin1, Rafael Orozco1, and Felix J. Herrmann1 1 Georgia Institute of Technology, * equal contribution},
  colorlinks=true,
  linkcolor={blue},
  filecolor={Maroon},
  citecolor={Blue},
  urlcolor={Blue},
  pdfcreator={LaTeX via pandoc}}

\makeatletter
\def\and{%
  \end{tabular}%
  \hskip 1em \@plus.17fil\relax
  \begin{tabular}[t]{c}}
\makeatother

\title{A Digital Twin for Geological Carbon Storage with Controlled
Injectivity}
\author{Abhinav Prakash Gahlot\textsuperscript{1*}, Haoyun
Li\textsuperscript{1*}, \and Ziyi Yin\textsuperscript{1}, Rafael
Orozco\textsuperscript{1}, and Felix J. Herrmann\textsuperscript{1} \\
\textsuperscript{1} Georgia Institute of Technology, \textsuperscript{*}
equal contribution}
\date{}

\begin{document}
\maketitle
\ifdefined\Shaded\renewenvironment{Shaded}{\begin{tcolorbox}[borderline west={3pt}{0pt}{shadecolor}, enhanced, breakable, interior hidden, frame hidden, sharp corners, boxrule=0pt]}{\end{tcolorbox}}\fi

\hypertarget{abstract}{%
\subsection*{Abstract}\label{abstract}}
\addcontentsline{toc}{subsection}{Abstract}

\emph{We present an uncertainty-aware Digital Twin (DT) for geologic
carbon storage (GCS), capable of handling multimodal time-lapse data and
controlling CO\textsubscript{2} injectivity to mitigate reservoir
fracturing risks. In GCS, DT represents virtual replicas of subsurface
systems that incorporate real-time data and advanced generative
Artificial Intelligence (genAI) techniques, including neural posterior
density estimation via simulation-based inference and sequential
Bayesian inference. These methods enable the effective monitoring and
control of CO\textsubscript{2} storage projects, addressing challenges
such as subsurface complexity, operational optimization, and risk
mitigation. By integrating diverse monitoring data, e.g., geophysical
well observations and imaged seismic, DT can bridge the gaps between
seemingly distinct fields like geophysics and reservoir engineering. In
addition, the recent advancements in genAI also facilitate DT with
principled uncertainty quantification. Through recursive training and
inference, DT utilizes simulated current state samples, e.g.,
CO\textsubscript{2} saturation, paired with corresponding geophysical
field observations to train its neural networks and enable posterior
sampling upon receiving new field data. However, it lacks
decision-making and control capabilities, which is necessary for full DT
functionality. This study aims to demonstrate how DT can inform
decision-making processes to prevent risks such as cap rock fracturing
during CO\textsubscript{2} storage operations.}

\hypertarget{introduction}{%
\subsection{Introduction}\label{introduction}}

Digital Twins refer to dynamic virtual replicas of subsurface systems,
integrating real-time data and employing advanced generative Artificial
Intelligence (genAI) methodologies, such as neural posterior density
estimation via simulation-based inference (Cranmer, Brehmer, and Louppe
2020) and sequential Bayesian inference (Papamakarios, Sterratt, and
Murray 2019; Kruse et al. 2021). Thanks to combination of these advanced
Bayesian inference techniques, our approach is capable of addressing
challenges of monitoring and controlling CO\textsubscript{2} storage
projects. These challenges include dealing with the subsurface's
complexity and heterogeneity (seismic and fluid-flow properties),
operations optimization, and risk mitigation, e.g.~via injection rate
control. Because our Digital Twin is capable of handling diverse
monitoring data, consisting of time-lapse seismic and data collected at
(monitoring) wells, it entails a technology that serves as a platform to
integrate seemingly disparate and siloed fields, e.g.~geophysics and
reservoir engineering. In addition, recent breakthroughs in genAI, allow
Digital Twins to capture uncertainty in a principled way (Yu et al.
2023; Herrmann 2023; Gahlot et al. 2023). By employing training and
inference recursively, the Digital Twin trains its neural networks on
samples of the simulated current state---i.e., the CO\textsubscript{2}
saturation/pressure, paired with simulated imaged seismic and/or data
collected at (monitoring) wells. These training pairs of the simulated
state and simulated observations are obtained by sampling the posterior
distribution,
\(\mathbf{x}_{k-1}\sim p(\mathbf{x}_{k-1}\vert \mathbf{y}^\mathrm{o}_{1:k-1})\),
at the previous timestep, \(k-1\), conditioned on field data,
\(\mathbf{y}^\mathrm{o}_{1:k-1}\), collected over all previous
timesteps, \(1:k-1\), followed by advancing the state to the current
timestep, followed by simulating (seismic/well) observations associated
with that state. Given these simulated state-observation pairs, the
Digital Twin's networks are trained, so they are current and ready to
produce samples of the posterior when the new field data comes
in---i.e.~\(\mathbf{x}_{k}\sim p(\mathbf{x}_{k}\vert \mathbf{y}^\mathrm{o}_{1:k})\).
While this new neural approach to data assimilation for
CO\textsubscript{2} storage projects provides what is called an
uncertainty-informed \emph{Digital Shadow}, it lacks decision making and
control, which would make it a Digital Twin (Thelen et al. 2022),
capable of optimizing storage operations while mitigating risks
including the risk of fracturing the cap rock by exceeding the fracture
pressure. The latter risk is illustrated in Figure 1, where the first
row contains simulated samples of the pressure difference at timestep
\(k=4\), between the reservoir pressure and hydraulic pressure, without
control. These samples for the simulated state exceed the fracture
pressure and are denoted by the red areas. This manuscript will
demonstrate how the Digital Twin can make informed decisions to avoid
exceeding the fracture pressure.

\hypertarget{methodology}{%
\subsection{Methodology}\label{methodology}}

To make uncertainty informed decisions on adapting the injection rate,
128 samples of the state (see second row Figure 1),
\(\{\mathbf{x}^{(m)}_{3}\}_{m=1}^{128}\), and permeability,
\(\{K^{(m)}\}_{m=1}^{128}\), are drawn at time-step \(k=3\) from the
posterior distribution,
\(\mathbf{x}_{3}\sim p(\mathbf{x}_{3}\vert \mathbf{y}^\mathrm{o}_{1:3})\),
for the state conditioned on the observed time-lapse data, and from the
distribution for the permeability, \(K\sim p(K)\). To find the optimized
injection rate, we first calculate for each sample, \(K\) and
\(\mathbf{x}^{(m)}_{3}\), the optimized injectivity by
\(\max_{q_3} q_3\Delta t \ \ \text{subject to} \ \ \mathbf{x}_{4}['p']<\mathbf{p}_{\max}\)
where \(\mathbf{p}_{\max}\) is the depth-dependent fracture pressure,
and \(\mathbf{x}_{4}=\mathcal{M}_3(\mathbf{x}_{3}, \mathbf{K}; q_3)\)
the state's time-advancement denoted by the symbol \(\mathcal{M}_3\).
For the fluid-flow simulations, the open-source Julia package
\href{https://github.com/slimgroup/JUDI.jl}{JUDI.jl} and
\href{https://github.com/sintefmath/JutulDarcy.jl}{JutulDarcy.jl} are
used. Results of these optimizations are included in Figure 2(a), which
contains a histogram of the empirical fracture frequency as a function
of injection rates at time \(k=3\). Given this histogram, our task is to
maximize the injection rate given a pre-defined confidence interval
(e.g.~95 \%), so that the fracture probability remains below a certain
percentage e.g.~1\%. With these simulations, and the fact that
fracture/no-fracture occurrences are distributed according to the
Bernoulli distribution, we will demonstrate that we are able to select
an injection rate that limits fracture occurrence to the prescribed
probability with a prescribed confidence interval. For instance, we can
compute \(Pr([\mathbf{x}_4['p']>p_{\max}]<0.01)<1-0.025\), which
corresponds to selecting an injection rate that leads to fracture rate
of \(<1\%\) with \(97.5\%\) confidence. The confidence interval is
halved, because only conservative (left) injection rates will be
selected (see Figure 2(b)).

\hypertarget{results}{%
\subsection{Results}\label{results}}

To calculate injection rates that mitigate the risk of exceeding the
fracture pressure, we proceed as follows. First, because the optimized
injection rates are close to the fracture pressure, we consider these
optimization as approximations to the injection rates where the fracture
pressure are exceeded. Next, Kernel Density Estimation (KDE) is applied
to produce the smooth red curve in Figure 2(a). This smoothed
probability function is used to calculate the Cumulative Density
Function (CDF), plotted in Figure 2(b). Using the fact that
non-fracture/fracture occurrence entails a Bernoulli distribution,
confidence intervals can be calculated,
\(\pm Z_{\frac{\alpha}{2}} \sqrt{\frac{\hat{p}(1 - \hat{p})}{128}}\)
where \(\hat{p}\) represents the CDF (blue line) and
\(Z_{\frac{\alpha}{2}}=1.96\) with \(\alpha=0.05\). From the CDF and
confidence intervals (denoted by the grey areas), the following
conclusions can be drawn: First, if the initial injection rate of
\(q_3 = 0.0500 m^3/s\) is kept, the fracture probability lies between
24.47 -- 40.71\% (vertical dashed line) and has a maximum likelihood of
32.59\%, which are all way too high. Second, if we want to limit the
fracture occurrence rate to 1\% (red dashed line), then we need to lower
the injection rate to \(q_3=0.0387\mathrm{m^3/s}\). To ensure the low
fracture occurrence rate of 1\%, the reduced injection rate is chosen as
the smallest injection rate within the confidence interval. As can be
observed from Figure 1 (third row) and Figure 2(b), lowering the
injection rate avoids exceeding the fracture pressure at the expense of
injecting less CO\textsubscript{2}. Out of 128 samples, 43 samples are
fractured with the initial injection rate, while only one sample is
fractured with the controlled injection rate.

\hypertarget{conclusion-and-discussion}{%
\subsection{Conclusion and discussion}\label{conclusion-and-discussion}}

We illustrate how Digital Twins can be used to mitigate risks associated
with CO\textsubscript{2} storage projects. Specifically, we used the
Digital Twin's capability to produce samples of its state (pressure),
conditioned on observed seismic and/or well data. Using these samples,
in conjunction with samples from the permeability distribution, we were
able to capture statistics on the fracture occurrence frequency as a
function of the injection rate. Given these statistics, we set a
fracture frequency and choose the corresponding injection rate as a
function of the confidence interval. By following this procedure,
exceeding the fracture pressure was avoided by lowering the injection
rate. The decision to lower the injection rate, and by which amount, was
informed by the Digital Twin, which uses seismic and/or well data to
capture reservoir's state including its uncertainty.

\begin{figure}

{\centering 

\includegraphics[width=0.9\textwidth,height=\textheight]{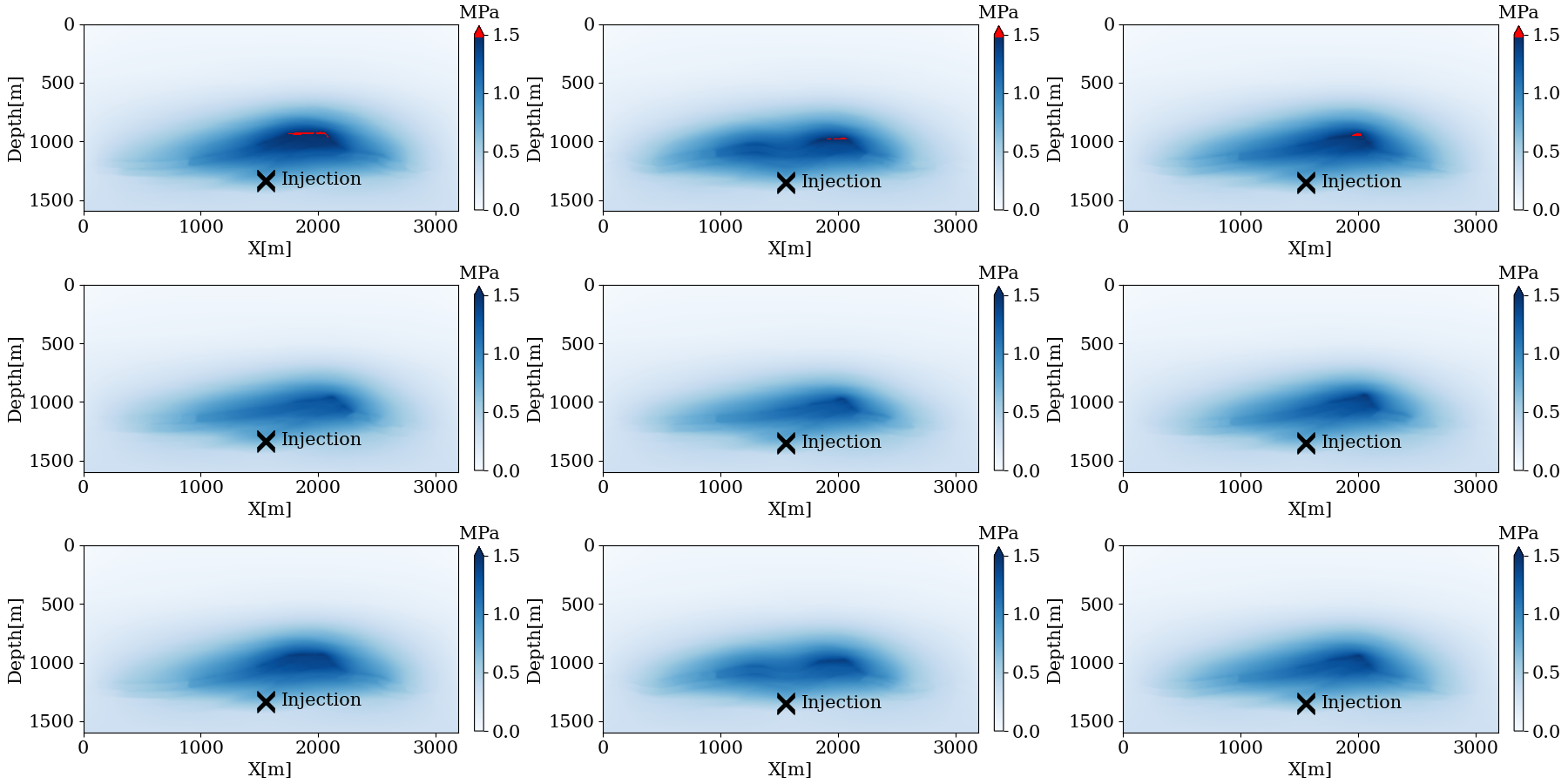}

}

\caption{\label{fig-flow}Pressure state at time-step \(k=3\) (second
row) and a comparative analysis of pressure outputs (first and third)
from the digital twin at time-step \(k=4\)}

\end{figure}

\begin{figure}

{\centering 

\includegraphics[width=0.9\textwidth,height=\textheight]{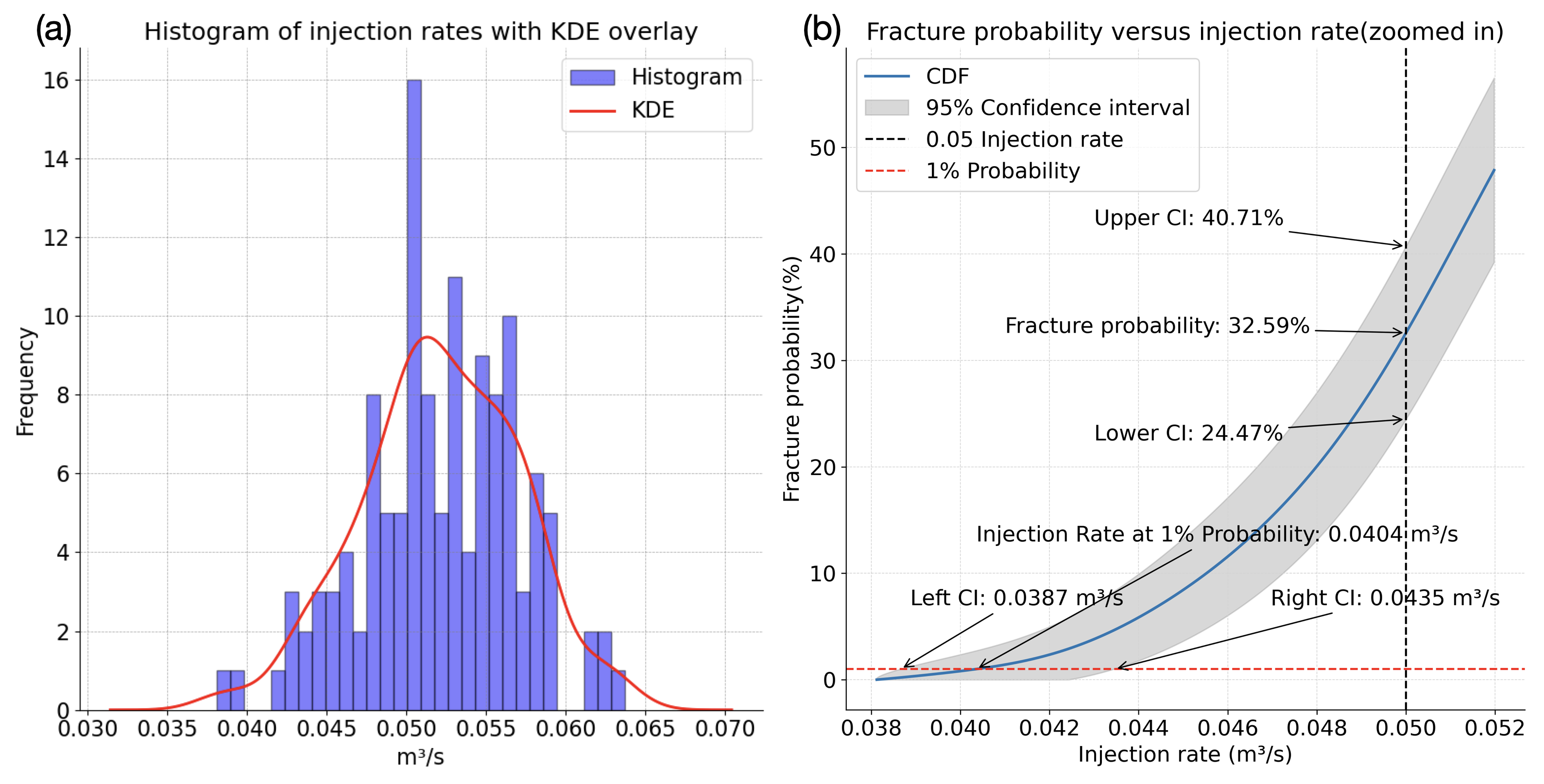}

}

\caption{\label{fig-fracture}Injection rate samples and KDE curve(a),
CDF curve of fracture probability versus injection rate with 95\%
confidence interval(b)}

\end{figure}

\hypertarget{references}{%
\subsection{References}\label{references}}

\hypertarget{refs}{}
\begin{CSLReferences}{1}{0}
\leavevmode\vadjust pre{\hypertarget{ref-doi:10.1073ux2fpnas.1912789117}{}}%
Cranmer, Kyle, Johann Brehmer, and Gilles Louppe. 2020. {``The Frontier
of Simulation-Based Inference.''} \emph{Proceedings of the National
Academy of Sciences} 117 (48): 30055--62.
\url{https://doi.org/10.1073/pnas.1912789117}.

\leavevmode\vadjust pre{\hypertarget{ref-gahlot2023NIPSWSifp}{}}%
Gahlot, Abhinav Prakash, Huseyin Tuna Erdinc, Rafael Orozco, Ziyi Yin,
and Felix J. Herrmann. 2023. {``Inference of CO2 Flow Patterns
{\textendash} a Feasibility Study.''}
\url{https://doi.org/10.48550/arXiv.2311.00290}.

\leavevmode\vadjust pre{\hypertarget{ref-herrmann2023president}{}}%
Herrmann, Felix J. 2023. {``President's Page: Digital Twins in the Era
of Generative AI.''} \emph{The Leading Edge} 42 (11): 730--32.

\leavevmode\vadjust pre{\hypertarget{ref-kruse2021hint}{}}%
Kruse, Jakob, Gianluca Detommaso, Ullrich Köthe, and Robert Scheichl.
2021. {``HINT: Hierarchical Invertible Neural Transport for Density
Estimation and Bayesian Inference.''} In \emph{Proceedings of the AAAI
Conference on Artificial Intelligence}, 35:8191--99. 9.

\leavevmode\vadjust pre{\hypertarget{ref-papamakarios2019sequential}{}}%
Papamakarios, George, David C. Sterratt, and Iain Murray. 2019.
{``Sequential Neural Likelihood: Fast Likelihood-Free Inference with
Autoregressive Flows.''} \url{https://arxiv.org/abs/1805.07226}.

\leavevmode\vadjust pre{\hypertarget{ref-thelen2022comprehensivea}{}}%
Thelen, Adam, Xiaoge Zhang, Olga Fink, Yan Lu, Sayan Ghosh, Byeng D
Youn, Michael D Todd, Sankaran Mahadevan, Chao Hu, and Zhen Hu. 2022.
{``A Comprehensive Review of Digital Twin---Part 1: Modeling and
Twinning Enabling Technologies.''} \emph{Structural and
Multidisciplinary Optimization} 65 (12): 354.

\leavevmode\vadjust pre{\hypertarget{ref-yu2023IMAGEmsc}{}}%
Yu, Ting-ying, Abhinav Prakash Gahlot, Rafael Orozco, Ziyi Yin, Mathias
Louboutin, and Felix J. Herrmann. 2023. {``Monitoring Subsurface CO2
Plumes with Sequential Bayesian Inference.''}
\url{https://slimgroup.github.io/IMAGE2023/SequentialBayes/abstract.html}.

\end{CSLReferences}

\end{document}